# Energy loss and damage production by heavy ions and strange quark matter in silicon

Sorina Lazanu
National Institute for Materials Physics,
POBox MG-7, Bucharest-Măgurele, Romania, e-mail: lazanu@infim.ro

Ionel Lazanu
University of Bucharest, Faculty of Physics, Department of Atomic and Nuclear Physics
POBox MG-11, Bucharest-Măgurele, Romania, e-mail: ionel.lazanu@gmail.com

Abstract

In this contribution the peculiarities of the behaviour of strange quark matter in respect to ordinary ions in silicon are investigated, and a tentative to identify possible observable effects of degradation is made.

## 1. Introduction

Strange quark matter is a hypothetical state of matter with unusual properties: stability, large mass range, low charge to mass ratio. The main objective of the present contribution is the investigation of the peculiarities of the behaviour of strange quark matter in respect to ordinary ions in silicon and the identification of possible observable effects of degradation. Starting from the predicted characteristics of the SQM, we consider the interaction processes of ordinary ions and SQM in silicon, putting into evidence those phenomena which produce irreversible bulk degradation in material and effects at the device level.

## 2. Short review on strange quark matter (SQM)

Strange quark matter is another state of matter [1, 2]. As it is known, quarks and gluons carry colour charges. The confinement character of the strong force prohibits the isolated existence of single quarks, but they can cluster in pairs or small groups, as mesons, baryons, or antibaryons. There is no basic physical principle known which excludes the existence of larger baryons. Strange quark matter is so-called because of the admixture of a large number of delocalized quarks (u,…, u, d,…, d, s,…, s). The mass range for this kind of matter may lie anywhere between the masses of light nuclei and that of neutron stars. SQM should be considered stable or relatively stable based on the following possible arguments [3 ÷ 5]:

- The (weak) decay of an s-quark from SQM into a d-quark would be suppressed or forbidden if the lowest single particle states are occupied.

- The strange quark mass is lower than the Fermi energy of the u or d quark in such a quark droplet; the opening of a new flavour degree of freedom tends to lower the Fermi energy and hence also the mass of the strangelet.
- SQM should have relatively small positive integer charge and the neutrality is realized by the existence of electrons, analogous to the case of atoms.

The unusual properties: stability, large mass range, low charge to mass ratio, low energy, are considered as characteristic for SQM. In particular, the low $Z/A$ ratio has been recognized as a crucial signature for the experimental identification of strangelets. If for ordinary nuclei the $Z/A$ ratio is usually between 0.33 ÷ 0.67, in colour flavour locked SQM (structures where quarks with different colour and quantum numbers form Cooper pairs), this is $\sim 0.3 \times A^{-1/3}$ and thus in ordinary strangelets it is approximately constant for small $A$, $A^{-2/3}$ for large $A$, reaching $A^{-1/3}$ only asymptotically.

If strange quark matter is absolutely stable or quasistable, the SQM produced in the early moments of the universe in the quark – hadron transition at about T ≈ 100 MeV survives the evolution and in principle could exist as heavy isotopes with an unusual high mass and low charge in terrestrial or stellar matter, in cosmic rays, or could be produced in collision experiments of heavy ions at high bombarding energies.

In experiments devoted to the study of primary CR nuclei, anomalous events have been observed [6 ÷ 8], with values of charge Z = 14 and mass numbers A = 350 and A = 450 respectively, with Z = 46 and A > 1000 and with Z = 20 and A = 460, and have been reported. Recently, two events were put in evidence in the AMS-01 data: $^{16}$He: Z = 2, A~16 and $^{54}$O: Z = 8, A = 54+8(−6) [9]. A short-lived strangelet candidate with mass 7.4 GeV and charge Z = −1 was separated by the NA52 experiment [10].

## 3. Effects in silicon due to heavy ions and SQM

The incident ion interacts with the electrons and with the nuclei of the semiconductor lattice. It loses its energy in several processes, which depend on the nature of the particle and on its energy. The effect of the interaction of the incident particle with the electrons of the target is ionisation. The nuclear interaction between the incident particle and the lattice nuclei produces bulk defects, depending on the nonionising energy loss (NIEL). The recoil nucleus (or nuclei in the case of inelastic processes) is displaced from the lattice site into interstitial positions. Then, the primary knock-on nucleus, if its energy is large enough, can produce the displacement of a new nucleus, and the process continues as long as the energy of the colliding nucleus is higher than the threshold for atomic displacements. The primary mechanism for defect formation during irradiation in semiconductors is the collision of the incoming particle with the atoms of the crystal, which leads to the departure of an atom to a rather large distance from its original site, i.e., to the formation of separated Frenkel pairs (vacancies and interstitials) as well as of the Si$_{FFCD}$ defect, that introduces a new type of symmetry in lattice [11, 12]:

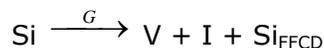

$$\text{Si} \xrightarrow{G} V + I + \text{Si}_{FFCD}$$

In silicon, the primary defects (V and I) are mobile over a broad temperature range, including room temperature, and diffusing they are trapped by other primary defects or by impurity atoms forming secondary defects or migrate to sinks. The main characteristics of the kinetics of defects are summarised in Ref. [12].

The knowledge of the dynamics of defects in silicon during and after irradiation permits the determination of the effects produced at the p-n junction detector level. Due to the low radiation flux of particles from background and to the effects induced by every ion, the main modification in the detector parameters is observed in the leakage current.

## 4. Results and discussions

The competition between nuclear and electronic stopping produced in a material by ions is a function on energy. It must be mentioned that at low incident ion kinetic energy, nuclear stopping is the dominant process and this behaviour depends on Z and A, so it is very different for hypothetical SQM in respect to ordinary ions. This behaviour could be observed in Figure 1.

The bulk degradation due to nuclear interactions, the NIEL, is represented in Figure 2 for a great number of ions and for all hypothetical candidates for SQM versus the kinetic energy of the ions

Also, the spatial distribution of the degradation simulated using the SRIM 2006-02 MC programme put in evidence major differences between ordinary ions and corresponding exotic cases, candidates for SQM.

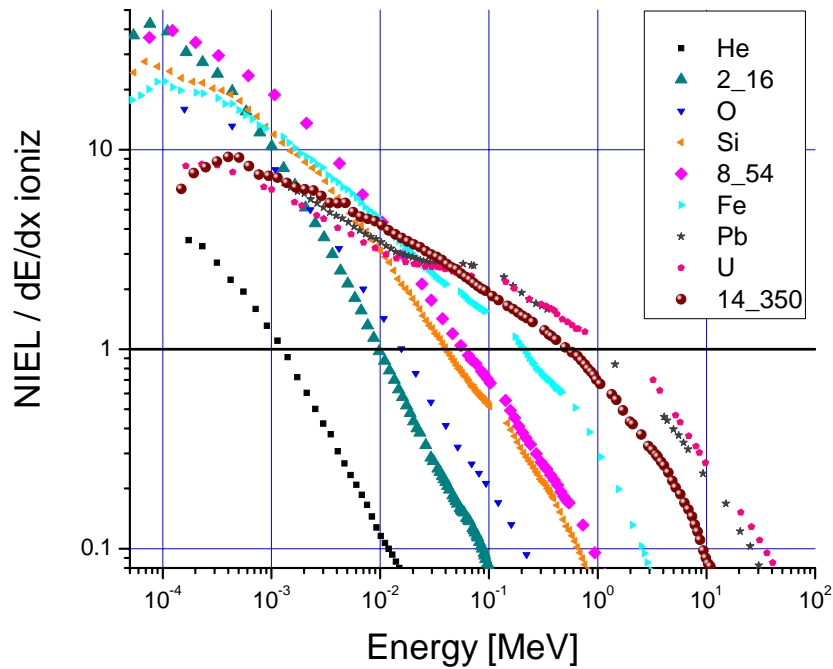

**Figure 1**.
Energy dependence of the ratio between NIEL and ionisation for ordinary ions and hypothetical SQM

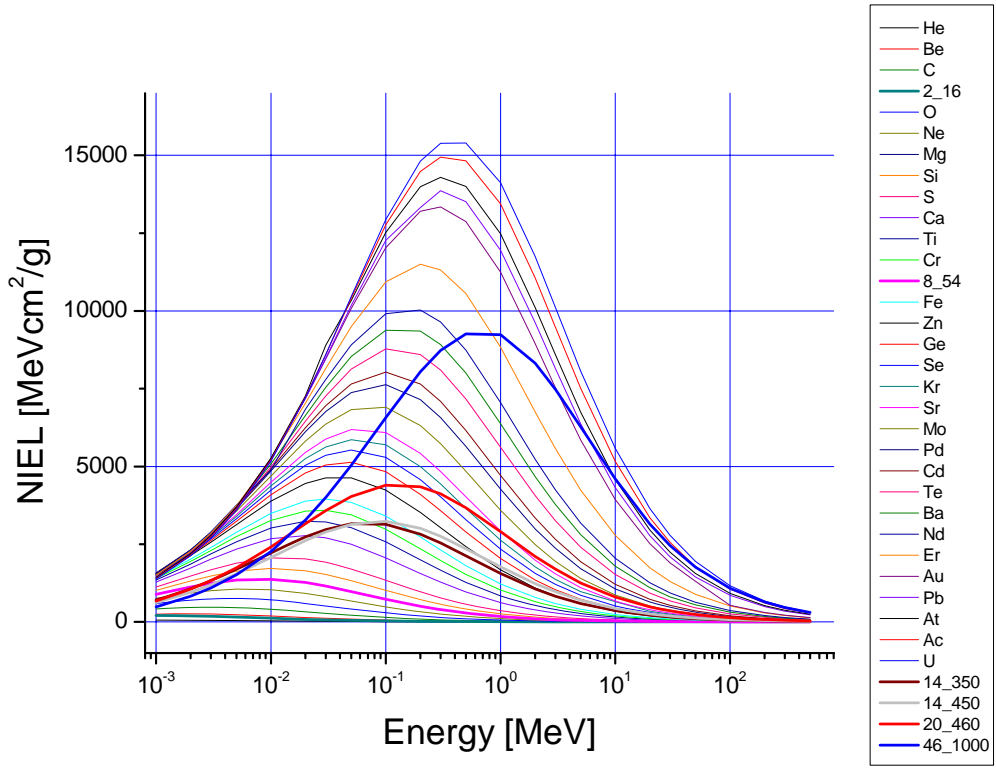

**Figure 2**.
Energy dependence of NIEL for ordinary ions and hypothetical candidates to SQM

In the calculations, analytical approximations [13] and MC simulations [14] have been used.

The effect at the device level is the increase of the density of the leakage current. So, for example, in a detector exposed for two years to a continuous flux of protons from cosmic rays, the passage of one SQM ion (Z=14, A=350) with energy 10 MeV produces an increase in the current density up to about $10^{-4}$ A/cm$^3$ from an initial value of $2 \times 10^{-9}$ A/cm$^3$.

## 5. Summary and possible conclusions

A very different behaviour of SQM ions was put in evidence in their interaction in silicon, in respect to ordinary ions. These aspects were observed in: NIEL, as a measure of bulk degradation, number of primary VI pairs per incident ion, characteristics of the interaction region. For the region where the incident ion deposits its energy, a significant increase of the density of the leakage current was estimated. More detailed studies are necessary to put in evidence aspects of experimental interest.

## Acknowledgments

This work was partially supported by the Romanian National University Research Council, under grant no. 28 /2007.